# PrivySense: Price Volatility based Sentiments Estimation from Financial News using Machine Learning

**Raeid Saqur***
Department of Computer Science
University of Toronto
raeidsaqur@cs.toronto.edu

**Nicole Langballe**
Department of Computer Science
University of Toronto
langballn@cs.toronto.edu

## Abstract

As machine learning ascends the peak of computer science zeitgeist, the usage and experimentation with sentiment analysis using various forms of textual data seems pervasive. The effect is especially pronounced in formulating securities trading strategies, due to a plethora of reasons including the relative ease of implementation and the abundance of academic research suggesting automated sentiment analysis can be productively used in trading strategies. The source data for such analyzers ranges a broad spectrum like social media feeds, micro-blogs, real-time news feeds, ex-post financial data etc. The abstract technique underlying these analyzers involve supervised learning of sentiment classification where the classifier is trained on annotated source corpus, and accuracy is measured by testing how well the classifiers generalizes on unseen test data from the corpus. Post training, and validation of fitted models, the classifiers are used to execute trading strategies, and the corresponding returns are compared with appropriate benchmark returns (for e.g., the S&P500 returns).

In this paper, we introduce a novel technique of using price volatilities to empirically determine the sentiment in news data, instead of the traditional reverse approach. We also perform meta sentiment analysis by evaluating the efficacy of existing sentiment classifiers and the precise definition of sentiment from securities trading context. We scrutinize the efficacy of using human-annotated sentiment classification and the tacit assumptions that introduces subjective bias in existing financial news sentiment classifiers.

## 1 Introduction

Sentiment analysis of opinion-rich (user-generated or professionally produced) textual data is a broad topic (e.g. product reviews, political blogs etc.). In this paper, we focus on its impact in the financial industry - more specifically, automatic sentiment estimation from financial news. Sentiment analysis is inevitably subjective and therefore varies based on the application and the bias of the reviewers. However, sentiment analysis for publicly-traded companies is a special case where the market serves as the judge [11]. The reaction to business news is reflected in the price change, along with other variables concerning price. Typically, sentiment analysis in financial economics refers to the deviation of market confidence indicators such as stock price and trading volumes [5]. This leads to a unique problem of defining sentiment in financial economics.

*I thank Prof. Tom McCurdy (Founder and Academic Director), and Eric Kang (Senior Research Associate) from Rotman Finance Lab at the University of Toronto, for their kind help with financial data access. All errors are mine.

Sentiment classification has been primarily researched outside of financial context. Sentiment outside financial domain is commonly inferred by comparing human-annotated sources or focusing on customer reviews such as movie reviews, product reviews and other publicly available sources. Inferring sentiment based on market reaction poses a new challenge as it is varies from the strong connection between text and reviews captured in previous work [1]. The definition of sentiment within financial news can be measured in several interpretations. The wording of the news can be analyzed with sentiment based classification used on reviews of products. However, sentiment can also be inferred through market reactions in terms of stock price, stock price volatility and trade volumes. All these are indicators of the subjective measure of sentiment. Research indicates that sentiment varies by domain and tuning sentiment analyzers to the task yields increased results [9].

In this paper, we argued that the definition of 'sentiment' in the financial domain should be precise and objective, and should be preferred over subjective human-annotated sentiment classifiers. We introduce a novel technique of empirically estimating news sentiment for financial news using stock price volatility instead of the reverse. To our knowledge, this was never attempted before, and has the potential to produce improved trading strategies and market returns.

## 2 Related Work

Previous research dating back over 40 years has demonstrated the connection between published news and market reaction [9]. Numerous studies have shown the published news affects the markets in profound ways, and impacts can be observed from changes in stock price, trade volumes, volatility, and even future firm earnings [5]. Research on the sentiment of financial news further explores this relationship to market reaction. Koppel and Shtrimberg (2006) established the relationship between the sentiment of news stories and price changes of a company stock . The impact of social media, and micro-blogging forums such as Twitter has been connected with stock returns correlating message volume and trading volume [15]. Studies have further explored this linked and predicted stock market returns from the sentiment on Twitter [2]. Investor opinions transmitted through social media have also been used to predict abnormal stock returns and earnings [4].

Using various forms of sentiment analysis for formulating trading strategies has also been researched. Zhang and Skiena (2010) used published news sentiment in a market-neutral trading strategy which provided favorable returns [16]. See Pand, Lee, et al. (2008) for detailed survey in this topic [13].

This paper builds off the work done by Kazemian, Zhao, and Penn (2016) which re-examined the tactical assumptions behind how sentiment analyzers are evaluated. In this paper, a trading strategy based on the sentiment of financial news was created and the underlying sentiment analyzer was examined based on market returns and classifier accuracy compared to human-annotated results. Results showed that the analyzer with the highest returns had the worst classification accuracy, showing the need for task-based performance measures. The paper stated that a more precise definition of news sentiment may lead to improved performance, suggesting the use of stock volatility as an approach.

## 3 Method and Materials

### 3.1 News Data

All news data used were retrieved from *Reuters* news data collection. First we created sets of 20 stocks randomly sampled from the universe of stocks that were listed and traded on either the NASDAQ or NYSE stock exchanges during our 'evaluation period': January 1, 2010 to October 30, 2017. The first set of stocks were roughly evenly weighted collection of 20 small-, mid-, and large-cap companies in the S&P500 list. The second set of 20 stocks were chosen from a list of ranked top 100 volatile stocks trading in NASDAQ [2]. The reasoning for choosing from multiple sets was to diversify our news training set corresponding to different levels of average volatility. To elucidate, with the presumption that stock prices in the S&P500 would fluctuate less to news compared to lower-cap stocks. Thus, the price volatility would be different for similar news for assets in different classes by market cap. Combining the two sampled sets calibrated the training set to eliminate this asymmetry. A news data crawler was implemented to retrieve all Reuters news document belonging to the stocks during the evaluation period. For our final data set, we coalesced the two sets and sampled 19 stocks with roughly even weights in terms of news document volumes and set origin. Trying to keep an even number of news documents for each ticker resulted in a significantly curtailed total number of documents (as more volatile and lower cap stocks had significantly lower number of news documents in Reuters data set). In the end, we had approximately 8000 news documents for the final set, which is comparable to the size used in the Kazemian paper [9], after discounting for duplicates.

[2] http://www.marketvolume.com/stocks/mostvolatile.asp

Finally, in order to test how well our trained model generalizes on randomly chosen industry news, we created a separate dataset consisting of only credit card companies during the evaluation period.

Please see Appendix A for details.

### 3.2 Stock Price Data and Volatility Calculation

We used the *Wharton Research Data Services* (WRDS) database for collecting all prices data for our chosen stock sets. The final news dataset was then joined with the prices dataset to add all price related pertinent columns to our final data (for e.g. 'prccd' (closing price)).

Price volatility calculation was an integral precursor for our experiments setup. We incorporated the '*news impact curve*' which measures how new information is incorporated into volatility estimates - outlined in the seminal paper by Engle and Ng (1993) into our volatility calculations. The autoregressive conditional heteroskedasticity (ARCH) model, originally introduced by Engle (1982) and its derivatives are the most popular class of parametric models for news impact curve. The pth order ARCH model, ARCH(p):

$$h_t = \omega + \sum_{i=1}^{P} \alpha_i \epsilon_{t-i}^2 \tag{1}$$

where $\alpha_1, ..., \alpha_p$, $\omega$ are constant parameters. The effect of a return shock i periods ago $(i <= p)$ on current volatility (at time $t$) is governed by $\alpha_i$. We expect that $\alpha_i < \alpha_j$ for $i > j$ - meaning, older news has lesser effect on current volatility. In ARCH(p) model, old news that arrived at the market more than p periods ago has no effect on current volatility.

In this paper, we have chosen to use a popular generalization of ARCH(p) by Bollerslev (1986) , the GARCH(p,q) model:

$$h_t = \omega + \sum_{i=1}^{P} \alpha_i \epsilon_{t-1}^2 + \sum_{i=1}^{q} \beta_i h_{t-i} \tag{2}$$

where $\alpha_1, ..., \alpha_p$, $\beta_1, ..., \beta_p$ $\omega$ are constant parameters. Due to its empirical success and wide adoption, we used the GARCH(1,1) model, where the effect of a return shock on current volatility declines geometrically over time.

Please see Appendix B for details.

## 4 Evaluations and Experiments

### 4.1 Feature Selection: Using TF-IDF, BM25

In order to complete sentiment analysis, feature selection for the collection of words in the news articles was compared across four methods. The first method was raw term frequency with a binary term frequency. The next method considered is count of term frequency, which linearly increases a word vectors weight based on frequency in a document. TF-IDF weight was another method implemented, TF-IDF is a statistical measure that weights the importance of a word in a document using term frequency and inverse document frequency [16]. TD-IDF increases the weight with rarity of term in the collection. The weight also increases with the frequency in the document but not linearly like with count of frequency.The TF-IDF formula implemented was:

$$w_i = tf_i \cdot idtf_i = tf_i \cdot log\frac{N}{df_i} \tag{3}$$

where $tf_i$ is the term frequency, $idtf_i$ is the inverse document frequency, and $N$ is the total number of documents in the collection [16].The last method used for feature weighting was BM25 is closely related to TF-IDF, which but has an added relevance score and is based on probabilistic approach [14]. The formula used for BM25 is:

$$w_i = tf_i \cdot idtf_i = \frac{(k_1 + 1) \cdot tf}{k_1((1 - b) + b \cdot \frac{dl}{avgdl}} \cdot log\frac{N - df + 0.5}{df + 0.5} \quad (4)$$

where $dl$ is the document length, $avgdl$ is the average document length in the collection, $k1$ and $b$ are parameters set to 1.2 and 0.95 respectively [16]. BM25 was implemented in this paper using bm25tf (term frequency) and bm25Idf (inverse term frequency) as outlined.

### 4.2 Feature Extraction using NLTK

The feature selection methods detailed in the preceding section had one serious limitation. They used bag-of-words representation of 'unigram' (or simplex) words, which has very limited ability to capture semantic meaning. We also noticed high-frequency important key words pervasive in financial documents appearing in both classes (positive and negative) of training data.

To elucidate, consider the simplex or unigram feature words: 'estimates', 'beat'. Now, consider the following sentences:

- Positive: "Quarterly revenue beat forecasted EPS estimates by $0.25"
- Negative: "The quarterly earnings per share failed to beat analyst estimates."

Here, bag-of-words features representation (even after applying TF-IDF and BM25 filtering) of features would include the words in both class labels. Thus, to circumvent this issue and improve our overall model, we explored avenues to extract features that capture semantic meaning.

Scrutinizing academic literature in the NLP domain, we found Su Nam Kim's (2010) paper on N-gram based Keyphrase extraction to capture semantic meaning most logical and applicable to our case.

In order to implement n-gram, we utilized the popular Python NLTK NLP library, which provides functions for word chunking, stemming, lemmatizing etc. The following Figure 1 shows a high-level pipeline architecture we implemented for n-gram based feature selection:

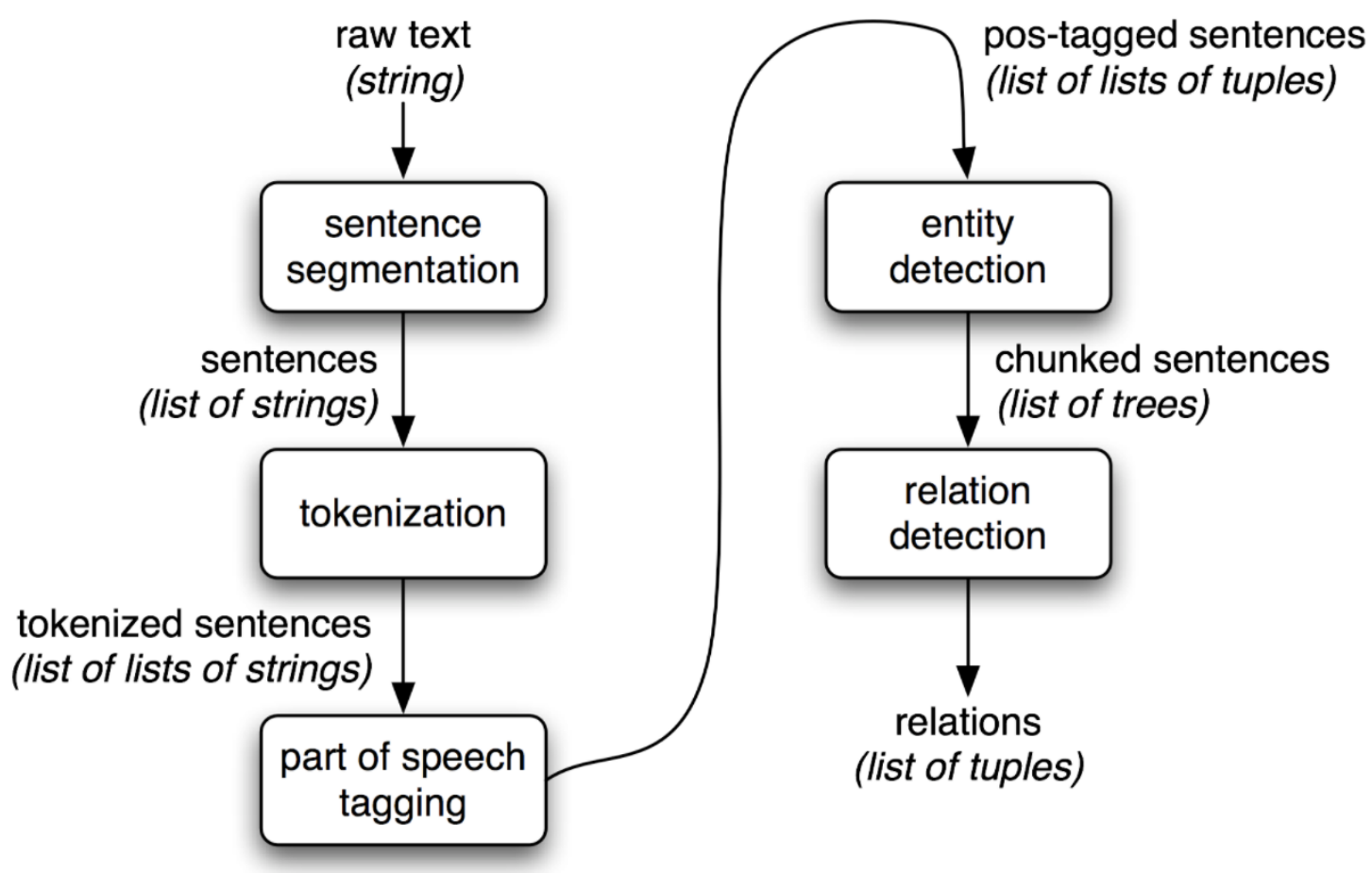

Figure 1: High level pipeline for feature extraction.

For part of speeching tagging to tokenized sentences, we had to define regular expression grammar rules. Figure 2 below shows the grammar definition we used for our experiments. We chose to use the same rules used by Su Nam et. al. [10], and did not attempt tuning rules for optimum results (falls under **future work** umbrella, discussed in later section). To exemplify, here we only included

**(Rule1)** NBAR = (NN⋆|JJ⋆)*(NN⋆)
e.g. *complexity, effective algorithm, distributed web-service discovery architecture*
**(Rule2)** NBAR IN NBAR
e.g. *quality of service, sensitivity of VOIP traffic, simplified instantiation of zebroid*

Figure 2: Regular expression grammar rules for feature selection.

nouns (NN) and adjectives (JJ) for chunks. However, including adverbs (RB) and/or verbs (VB) in our grammar rules could plausibly improve semantic value captured by chunks.

Additionally, we performed word normalization of all features by using 'stemmer' (PorterStemmer) to stem or convert words to root words, and then 'lemmatizer'(WordNet) to convert root words to valid dictionary words - thus ensuring all similar words account for the correct frequency count. For e.g. 'performed', 'performing', 'perform' are treated as one normalized word 'perform' with a frequency count of 3 instead of three different features with count of 1.

Performing n-gram based feature extraction also allowed us to logically reduce feature dimensions (by filtering out noisy words). Post extraction, we removed all unigram word features with less than frequency count of 3 from our features set, thus ensuring our features set contains n-gram terms and simplex or unigram words with minimum set frequency.

*Note*: we also did not tune the 'n' parameter, i.e. the word sequence count in our n-gram features. We used minimum '2' and maximum '5' for our sequence length, but it's plausible to optimize the results by tuning this parameter (falls under **future work** umbrella, discussed in later section). Johannes (1998) [8] posits that using word sequences of length 2 or 3 are most useful for Reuters news articles, and higher grams show negligible performance improvement, so we don't expect significant performance boost by tuning this parameter.

Running our classifier with preliminary n-gram model immediately performed better than fully tuned feature extraction models implemented in preceding section (please see 5), with illustrative performance improvement.

### 4.3 Sentiment Analysis and Intrinsic Evaluation

To validate our approach against previous work completed by Kazemian, Zhao, and Penn (2016) and Pang and Lee (2004) our sentiment analyzer was examined on the movie review data set. Pang and Lee's (2004) movie reviews dataset contains 1000 positive and 1000 negative reviews. Data preprocessing, feature selection and machine learning methods outlined in Section 3.1 and Section 4.1 were implemented on the movie reviews dataset as a sanity check. Data was prepossessed to remove stop words, punctuation and, words with a length of one. An additional processing step was completed to remove all words with a frequency of 3 or less in the training corpus. The feature selection methods mentioned in Section 4.1 were employed with machine learning methods Naive Bayes and Support Vector Machines with a linear kernel. The hyper-parameters for each model were tuned with Grid Search and 10 fold cross validation. The optimal parameters found were an alpha parameter of 1 for Naive Bayes and a regularization parameter of 1 for SVM using L2 penalty and hinge loss. The classification accuracy reported in the Figure 3 is the average for model over 10 folds. Stratified k-fold cross validation was used in order to preserve the percentage of a class in each fold, to prevent bias in a sample.

The highest accuracy obtained was 87.4% using TF-IDF features and Support Vector Machines as shown in Figure 3. Term Presence and BM25 features with Support Vector Machines also had a high performance on the movie reviews dataset. The results obtained are comparable with Kazemian, Zhao, and Penn (2016) accuracy of 86.85% and Pang and Lee (2004) accuracy of 86.4%. This shows our our sentiment analyzer algorithm is a comparable approach to the previous work we are building on. This intrinsic evaluation validates our approach to training a sentiment analyzer using Support Vector Machines and Feature Selection.

| Feature Selection Method | Linear SVM | Naïve Bayes |
|---|---|---|
| Term Presence | 85.3% | 79.5% |
| Raw Term Frequency | 83.1% | 79.5% |
| TF-IDF Features | 87.4% | 79.4% |
| BM25 Features | 85.7% | 79.4% |

Figure 3: Average 10 fold cross validation accuracy on movie reviews dataset

## 5 Demonstration

The sentiment analyzer was trained and tested on our combined final dataset of 19 stocks to explore the relationship between the release of news and stock volatility. We also ran a generalization test on the 'credit card companies' data set as outlined in Section 3.

Feature selection methods referenced in Section 4.1 and Section 4.2 were employed with machine learning techniques Naive Bayes and Support Vector Machines using a linear kernel. The sentiment classifiers were evaluated on training and test data separated with an 80-20 split. For the initial evaluation, price volatility was calculated using closing stock price on the news release date and one day following (using equation: $ln(\frac{P_t}{P_{t-1}})$). The sentiment of news was classified based on the price volatility, with a positive value corresponding to a positive label and a negative value corresponding to a negative label. In the initial experiment, zero was used as the threshold for negative and positive.

Figure 4 and Figure 5 shows the performance of the sentiment classifiers on the test sets. Feature selection with NLTK had the highest accuracy and F1 score on the both datasets. Linear support vector machines slightly outperformed the Naive Bayes classifier on the S&P500 dataset with an F1 score of 0.62 versus 0.60, the results for credit card company data had an even closer margin.

| | 19 Stocks from S&P500 | | Credit Card Companies | |
|---|---|---|---|---|
| Feature Selection Method | Linear SVM | Naïve Bayes | Linear SVM | Naïve Bayes |
| Term Presence | 60.5% | 58.4% | 61.6% | 59.2% |
| Raw Term Frequency | 61.4% | 58.4% | 58.4% | 59.2% |
| TF-IDF Features | 59.0% | 58.4% | 59.2% | 60.8% |
| BM25 Features | 60.0% | 58.4% | 60.8% | 60.8% |
| Feature Extraction using NLTK | 62.9% | 62.9% | 63.7% | 63.9% |

Figure 4: Test Accuracy (%) for the sentiment classifiers

| | 19 Stocks from S&P500 | | Credit Card Companies | |
|---|---|---|---|---|
| Feature Selection Method | Linear SVM | Naïve Bayes | Linear SVM | Naïve Bayes |
| Term Presence | 0.61 | 0.58 | 0.61 | 0.61 |
| Raw Term Frequency | 0.61 | 0.58 | 0.57 | 0.61 |
| TF-IDF Features | 0.58 | 0.58 | 0.58 | 0.61 |
| BM25 Features | 0.59 | 0.58 | 0.60 | 0.61 |
| Feature Extraction using NLTK | 0.62 | 0.60 | 0.63 | 0.64 |

Figure 5: F1 Score for the sentiment classifiers

To further analyze the performance of the sentiment classifiers positive and negative F1 scores are examined in Figure 6. Feature selection with NLTK preformed the highest in terms of overall accuracy and F1 score on both datasets, the improvement is attributed to an increase in the classification of positive volatility and a slight decrease in the negative classification.

| | 19 Stocks from S&P500 | | | | Credit Card Companies | | | |
|---|---|---|---|---|---|---|---|---|
| | Linear SVM | | Naïve Bayes | | Linear SVM | | Naïve Bayes | |
| Feature Selection Method | Postivie F1 Score | Negative F1 Score | Postivie F1 Score | Negative F1 Score | Postivie F1 Score | Negative F1 Score | Postivie F1 Score | Negative F1 Score |
| Term Presence | 0.63 | 0.58 | 0.63 | 0.53 | 0.67 | 0.55 | 0.63 | 0.58 |
| Raw Term Frequency | 0.64 | 0.59 | 0.63 | 0.53 | 0.65 | 0.49 | 0.63 | 0.58 |
| TF-IDF Features | 0.65 | 0.51 | 0.63 | 0.53 | 0.65 | 0.51 | 0.63 | 0.58 |
| BM25 Features | 0.66 | 0.52 | 0.63 | 0.53 | 0.67 | 0.52 | 0.63 | 0.58 |
| Feature Extraction using NLTK | 0.71 | 0.47 | 0.73 | 0.42 | 0.72 | 0.49 | 0.70 | 0.54 |

Figure 6: Positive and Negative F1 scores for the sentiment classifiers

Figure 7 displays the confusion matrix for the top performing classifier, the support vector machines using feature selection with NLTK. The classifier has a higher accuracy when predicting the positive class.

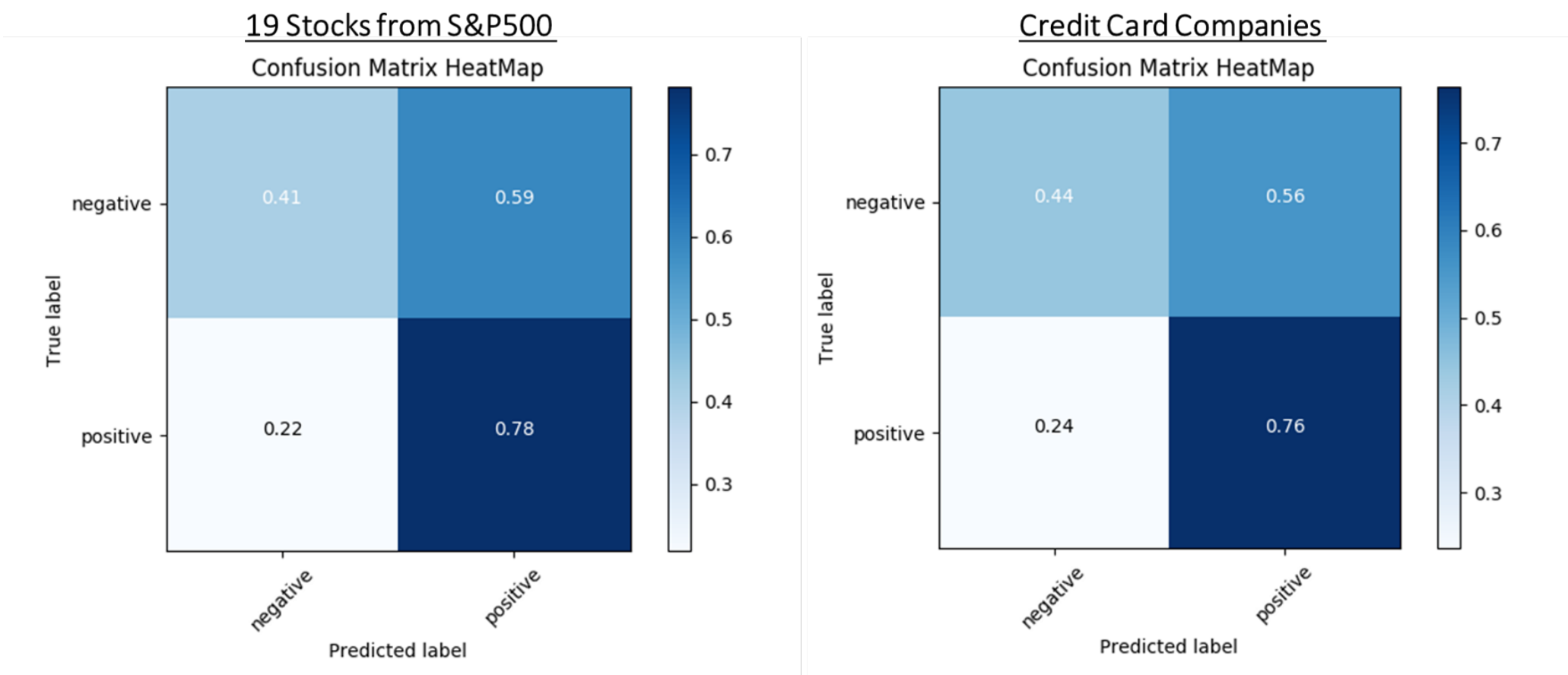

Figure 7: Normalized confusion matrix with SVM classification and Feature Selection using NLTK

The final experiment was to tune the price volatility formula with the most accurate classifier found in the initial experiment. This was the SVM classifier using feature selection with NLTK on the credit card company dataset. Stock price volatility was calculated using the formula outlined in Section 3.2. Figure 13 in Appendix B.2 displays the test accuracy for the price volatility calculation with the parameter p between 0 and 5. P equal to 0 is the baseline condition and price volatility calculation used in the initial experiment. The optimal value of p is found to be equal to 3, providing improved test accuracy over the baseline from the initial experiment. Please see Figure 14 in Appendix B.2 for complete 'p' parameter tuning results. The following Figure 8 shows the performance of the tuned model:

| Test Error (%) | F1 Score | Postivie F1 Score | Negative F1 Score |
|---|---|---|---|
| 71.1% | 0.70 | 0.78 | 0.57 |

Figure 8: Test results for the SVM classifier using NLTK n-gram features selection and p=3 for price volatility formula on the credit card data set

The result of the SVM classifier with p tuned to 3 is displayed in 8. This outperformed all sentiment classifiers in the initial experiment by a margin of 7.2%, the F1 score was improved by 0.6. The classifier had an improved positive F1 score from 0.72 to 0.78 and the negative F1 score increased from 0.49 to 0.57 with the tuning of the price volatility formula. The improved negative F1 score improves the classification of "negative" results by a significant margin. As can be seen in Figure 9, the classifier has better precision in predicting "positive" outcomes, a result consistent throughout the experiment.

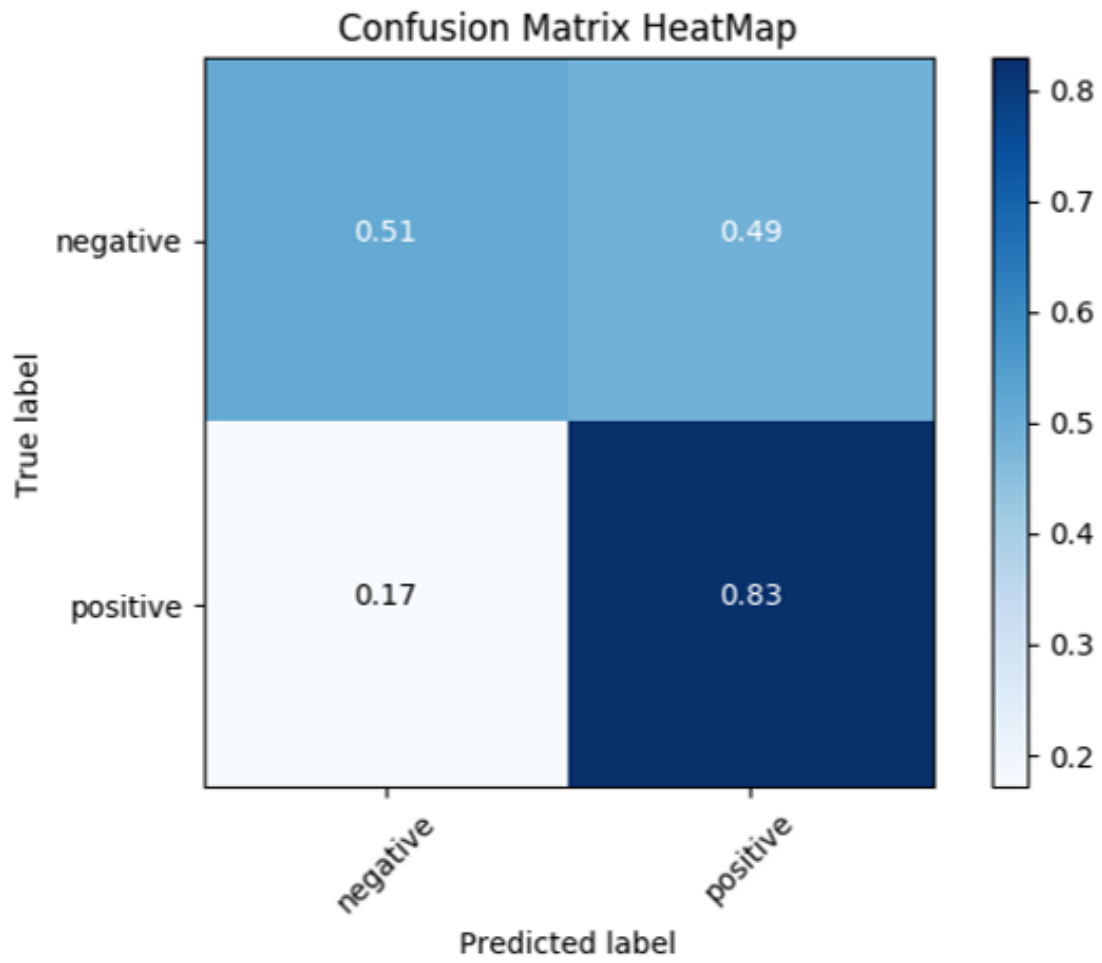

Figure 9: Normalized Confusion Matrix the SVM classifier using feature selection NLTK and p=3 for price volatility formula on the credit card data set

# 6 Limitations and Future Work

## 6.1 Limitations and Challenges

• **Small training data set**: a limiting factor was the size of the dataset, 16,256 news articles contained approximately 11,000 unique words. Providing accurate sentiment analysis prediction on future articles requires a move in-depth understanding of language and a significantly larger training corpus.

• **Impact attribution of news data on price volatility**: news data was extracted from Reuters, however we know that not every news story published will occur in Reuters. Therefore, various other sources that generate textual news data (like other news publications - e.g. Financial Post, financial report filings, analyst reports, social media posts - e.g. Twitter, Facebook etc.) were ignored in our experiments. Thus, by training our model using only one source of data is sure to introduce some attribution bias. However, research modeling entails simplification, and using Reuters only data set, in our opinion, is an acceptable proxy - as also validated by numerous other academic literature that uses this same technique.

• **Impact attribution of exogenous market factors on price volatility** : This is arguably the biggest limitation of our experiments, and perhaps a major weakness of our technique overall. Stock price volatility of an individual stock is not limited to news endogenous to itself, but multitude of exogenous market factors: like industry specific events, overall macro- and socio-economic events. However, it is our belief that such exogenous events, if pertinent enough to induce price volatility, would be covered in news data (may be not in news specific to a company perhaps, but as part of overall industry news). So, we can address that issue by refining news extraction and features selection techniques.

## 6.2 Future Extensions

• **Evaluation of trading strategies using 'PrivySense'**: due to imposed space constraint and time limitation, we considered a detailed 'task-based evaluation' of PrivySense beyond the scope for the first submission of this paper. However, we consider this to be imperative for the next version, as it would allow us to empirically compare our model's returns with other literature, including the Kazemian's paper where the model with best market returns had 3.46% 30-day returns with intrinsic (sentiment) classification accuracy of approximately 62% (much lower than PrivySense) [9].

• **Parametric tuning and optimization**: we have outlined the possibility of overall model performance improvement in a few sections - where we haven't tuned hyper-parameters. Most notably, for our price volatility calculations, we did not tune and fit the GARCH(p, q) parameters for the 'news impact curve'. Also, the Engle and Ng (1993) paper offers other models (GJR and EGARCH) that capture the news impact asymmetry of positive and negative news classes as better performers than our simple GARCH(1,1) model. Thus, we would like to optimize our model selection in this regard.

We would also like to explore, incorporate and optimize advanced NLP techniques to better capture semantic meaning from news data. For e.g., tune the grammar rules used for part of speech tagging and word chunking. We expect high accuracy gains in this aspect as financial documents have domain specific linguistic uniqueness in structure. For e.g., consider the linguistic similarity in quarterly reports news or SEC filings across the S&P universe companies.

• **Address limitations**: in the future, we would like to fix some of the limitations outlined in this paper. For e.g., we would like to increase the size of our curated dataset for training our model. This can be easily achieved by using our news crawler already implemented. For e.g., we would like to train our classifiers using news documents of all NASDAQ listed companies and evaluate its performance. Such large datasets would also increase the computation time significantly - and we plan to address that by using distributed GPUs as our computation machine.

## 7 Conclusion

**PrivySense** is a novel approach to evaluate sentiment in financial textual data. Using the proposed technique, we infer sentiment in news data using price fluctuations of asset prices. We trained a supervised support-vector-machine model that achieved 71.1% accuracy on generalized data set.

The idea for this experiment came from Kazemian, Zhao, and Penn's (2016) paper which examined the underlying sentiment analyzer in a trading strategy, finding the analyzer that performed highest on market returns performed lowest on traditional sentiment score. The results of the paper has sentiment classifiers performing between 62.09% and 81.14% when evaluated on human-subject annotations [9]. The classification accuracy achieved in this experiment of 71.1% is within the range of this study.

The disconnect between traditional sentiment and market reaction was further explored by evaluating the news corpus with a high-quality traditional sentiment classifier. The Microsoft Azure Text Analytics API was tested on the news data corpus to compare sentiment and stock price volatility. The Microsoft text analytics sentiment score for the Reuters news corpus from 2015-2017 containing randomly selected stocks from the S&P500 was compared against binary price volatility calculated 1-day after news release. The results were 52% accuracy for correctly predicting price volatility, with positive sentiment corresponding to a positive price volatility and negative sentiment corresponding to a negative price volatility. This result further illustrates the complex relationship between news sentiment and market reaction. The failure of Microsoft API to explain price volatility shows the need for exploring newer approaches to financial data sentiment analysis.

In this experiment, we have explored a new approach by establishing a relationship between price volatility and the release of financial news directly gauging the market reaction with a task based sentiment classifier. Future work would be to further evaluate the results achieved by utilizing this sentiment analyzer in a trading strategy.

# Appendices

## A Methods and Materials

### A.1 News Data

The following Figure 10 shows a sample snippet of news documents mined from Reuters news data collection.

| ticker | exchange | timestamp | story_headline | story_full | story_rank |
|---|---|---|---|---|---|
| MSFT | NASDAQ | 20171017 | RPT-INSIGHT-Microsoft responded quietly after detecting secret database hack in 2013 | Oct 17 Microsoft Corp's secret internal database for tracking bugs in its own software was broken into by a highly sophisticated hacking group more than | normal |
| MSFT | NASDAQ | 20171017 | INSIGHT-Microsoft responded quietly after detecting secret database hack in 2013 | Oct 17 Microsoft Corp's secret internal database for tracking bugs in its own software was broken into by a highly sophisticated hacking group more than | normal |
| MSFT | NASDAQ | 20171016 | BRIEF-Microsoft nominates two new members to its board of directors$— | * Microsoft - $—announced two nominations to its board of directors$— | topStory |
| MSFT | NASDAQ | 20171016 | BRIEF-Microsoft says CEO Satya Nadella's 2017 total compensation was $20.0 million $—$— | * Microsoft says CEO Satya Nadella's FY 2017 total compensation was $20.0 million versus $17.7 million FY 2016 - SEC filing$—$— | normal |
| MSFT | NASDAQ | 20171016 | U.S. Supreme Court to decide major Microsoft email privacy fight | WASHINGTON The U.S. Supreme Court on Monday agreed to resolve a major privacy dispute between the Justice Department and Microsoft Corp over w | normal |
| MSFT | NASDAQ | 20171016 | UPDATE 3-U.S. Supreme Court to decide major Microsoft email privacy fight | * Prosecutors worry about a ruling harming criminal probes (Adds reaction from digital rights advocacy group Justice Department declining comment) | normal |
| MSFT | NASDAQ | 20171016 | U.S. Supreme Court to decide Microsoft email privacy dispute | WASHINGTON Oct 16 The U.S. Supreme Court on Monday agreed to resolve a major privacy dispute between the Justice Department and Microsoft Corp | normal |
| MSFT | NASDAQ | 20171013 | BRIEF-Bit Evil signs non-disclosure agreement with Microsoft Corp | * SAID ON THURSDAY THAT IT HAS SIGNED A NON-DISCLOSURE AGREEMENT WITH MICROSOFT CORPORATION | topStory |
| MSFT | NASDAQ | 20171012 | BRIEF-Microsoft Amazon.com announce Gluon making deep learning accessible to all developers | * Microsoft - Amazon.com and Microsoft announce Gluon making deep learning accessible to all developers Source text for Eikon: Further company cove | topStory |
| MSFT | NASDAQ | 20171012 | BRIEF-Microsoft says Wipro to host several business-critical enterprise applications on Azure$— | * Says Wipro to host several business-critical enterprise applications on Azure$— Source text: [Microsoft today announced that Wipro Limited (NYSE: WI | normal |
| MSFT | NASDAQ | 20171011 | Microsoft Apple among companies urging U.S. Supreme Court to weigh gay workers' rights | Dozens of companies including Alphabet Inc$—Es Google Apple Inc Microsoft Corp and Viacom Inc have asked the U.S. Supreme Court to address wheth | topStory |
| MSFT | NASDAQ | 20171011 | Microsoft Apple among companies urging U.S. Supreme Court to weigh gay workers' rights | Oct 11 Dozens of companies including Alphabet Inc$—Es Google Apple Inc Microsoft Corp and Viacom Inc have asked the U.S. Supreme Court to addres | normal |
| MSFT | NASDAQ | 20171011 | Exclusive: Despite sanctions Russian organizations acquire Microsoft software | By Gleb Stolyarov Anastasia Teterevleva and Anastasia Lyrchikova | normal |
| MSFT | NASDAQ | 20171011 | Factbox: Entities subject to sanctions that bought Microsoft products | Software produced by Microsoft Corp has been acquired by entities in Russia and Crimea that are subject to sanctions barring companies based in the Uni | normal |
| MSFT | NASDAQ | 20171010 | Microsoft looks at whether Russians bought U.S. ads on search engine | SAN FRANCISCO Microsoft Corp said on Monday it was looking into whether Russians bought U.S. election ads on its Bing search engine or on other Micro | topStory |
| MSFT | NASDAQ | 20171010 | UPDATE 1-Microsoft looks at whether Russians bought U.S. ads on search engine | SAN FRANCISCO Oct 9 Microsoft Corp said on Monday it was looking into whether Russians bought U.S. election ads on its Bing search engine or on other | normal |
| MSFT | NASDAQ | 20171009 | Microsoft looks at whether Russians bought U.S. ads on search engine | SAN FRANCISCO Oct 9 Microsoft Corp said on Monday it was looking into whether Russians bought U.S. election ads on its Bing search engine or on other | topStory |
| MSFT | NASDAQ | 20171003 | MessageBird in biggest early-stage funding for European software firm | AMSTERDAM/LONDON Oct 3 Dutch start-up messaging company MessageBird has landed $60 million in first-round funding the largest ever early-stage | normal |
| MSFT | NASDAQ | 20171002 | Microsoft to end Groove streaming service; offers Spotify migration | Microsoft Corp said on Monday it would discontinue its Groove Music Pass subscription service allowing its existing customers to move their playlists and | topStory |
| MSFT | NASDAQ | 20171002 | Microsoft to end Groove streaming service; offers Spotify migration | Oct 2 Microsoft Corp said on Monday it would discontinue its Groove Music Pass subscription service allowing its existing customers to move their playlis | normal |
| MSFT | NASDAQ | 20170929 | Reuters hosts Newsmakers with Microsoft CEO Satya Nadella and RBC President and CEO David McKay | This week Reuters hosted two Newsmaker events featuring Microsoft CEO Satya Nadella and RBC President and CEO David McKay. | topStory |
| MSFT | NASDAQ | 20170928 | Microsoft search engine Bing to focus on PC search market: CEO | NEW YORK Microsoft Corp Chief Executive Officer Satya Nadella said on Wednesday the company's search engine Bing will focus on expanding in the PC | topStory |
| MSFT | NASDAQ | 20170928 | BRIEF-Microsoft collaborates with EY to launch EY synapse automotive$— | * Ey-Announces that it is collaborating with Microsoft to launch EY Synapse automotive a broad data analytics solution for automotive industry$— Sourc | normal |
| MSFT | NASDAQ | 20170927 | UPDATE 1-Microsoft search engine Bing to focus on PC search market -CEO | NEW YORK Sept 27 Microsoft Corp Chief Executive Officer Satya Nadella said on Wednesday the company's search engine Bing will focus on expanding i | topStory |
| MSFT | NASDAQ | 20170927 | Microsoft search engine Bing to focus on PC search market -CEO | NEW YORK Sept 27 Microsoft Corp Chief Executive Officer Satya Nadella said on Wednesday the company's search engine Bing will focus on expanding i | normal |
| MSFT | NASDAQ | 20170927 | LIVE REUTERS NEWSMAKER - Interview with Microsoft CEO Satya Nadella | Sept 27 Microsoft CEO Satya Nadella will speak to Reuters about artificial intelligence virtual reality and quantum computing. Tune in now to watch live | normal |
| MSFT | NASDAQ | 20170927 | LIVE REUTERS NEWSMAKER - Interview with Microsoft CEO Satya Nadella | Sept 27 Microsoft CEO Satya Nadella will speak to Reuters about artificial intelligence virtual reality and quantum computing. | normal |

Figure 10: Sample news document mined from Reuters news data.

### A.2 Prices Data

The following Figure 11 shows a sample snippet of prices csv file from WRDS research services database.

| gvkey | iid | datadate | tic | cusip | conm | prccd | prccd_vol | prchd | prcld | prcod | prcstd | trfd | exchg | secstat | tpci | cik | fic |
|---|---|---|---|---|---|---|---|---|---|---|---|---|---|---|---|---|---|
| 1690 | 1 | 20091102 | AAPL | 37833100 | APPLE INC | $ 189.31 | | 192.88 | 185.57 | 189.8 | 3 | 1.09566295 | 14 | A | 0 | 320193 | USA |
| 1690 | 1 | 20091103 | AAPL | 37833100 | APPLE INC | $ 188.75 | -0.30% | 189.52 | 185.92 | 187.85 | 3 | 1.09566295 | 14 | A | 0 | 320193 | USA |
| 1690 | 1 | 20091104 | AAPL | 37833100 | APPLE INC | $ 190.81 | 1.09% | 193.85 | 190.23 | 190.73 | 3 | 1.09566295 | 14 | A | 0 | 320193 | USA |
| 1690 | 1 | 20091105 | AAPL | 37833100 | APPLE INC | $ 194.03 | 1.67% | 195 | 191.82 | 192.4 | 3 | 1.09566295 | 14 | A | 0 | 320193 | USA |
| 1690 | 1 | 20091106 | AAPL | 37833100 | APPLE INC | $ 194.34 | 0.16% | 195.19 | 192.4 | 192.51 | 3 | 1.09566295 | 14 | A | 0 | 320193 | USA |
| 1690 | 1 | 20091109 | AAPL | 37833100 | APPLE INC | $ 201.46 | 3.60% | 201.9 | 196.26 | 196.94 | 3 | 1.09566295 | 14 | A | 0 | 320193 | USA |
| 1690 | 1 | 20091110 | AAPL | 37833100 | APPLE INC | $ 202.98 | 0.75% | 204.98 | 201.01 | 201.02 | 3 | 1.09566295 | 14 | A | 0 | 320193 | USA |
| 1690 | 1 | 20091111 | AAPL | 37833100 | APPLE INC | $ 203.25 | 0.13% | 205 | 201.83 | 204.56 | 3 | 1.09566295 | 14 | A | 0 | 320193 | USA |
| 1690 | 1 | 20091112 | AAPL | 37833100 | APPLE INC | $ 201.99 | -0.62% | 204.87 | 201.43 | 203.14 | 3 | 1.09566295 | 14 | A | 0 | 320193 | USA |
| 1690 | 1 | 20091113 | AAPL | 37833100 | APPLE INC | $ 204.45 | 1.21% | 204.83 | 202.07 | 202.87 | 3 | 1.09566295 | 14 | A | 0 | 320193 | USA |
| 1690 | 1 | 20091116 | AAPL | 37833100 | APPLE INC | $ 206.63 | 1.06% | 208 | 205.01 | 205.48 | 3 | 1.09566295 | 14 | A | 0 | 320193 | USA |
| 1690 | 1 | 20091117 | AAPL | 37833100 | APPLE INC | $ 207.00 | 0.18% | 207.44 | 205 | 206.08 | 3 | 1.09566295 | 14 | A | 0 | 320193 | USA |
| 1690 | 1 | 20091118 | AAPL | 37833100 | APPLE INC | $ 205.96 | -0.50% | 207 | 204 | 206.54 | 3 | 1.09566295 | 14 | A | 0 | 320193 | USA |
| 1690 | 1 | 20091119 | AAPL | 37833100 | APPLE INC | $ 200.51 | -2.68% | 204.61 | 199.8 | 204.61 | 3 | 1.09566295 | 14 | A | 0 | 320193 | USA |
| 1690 | 1 | 20091120 | AAPL | 37833100 | APPLE INC | $ 199.92 | -0.29% | 200.39 | 197.76 | 199.15 | 3 | 1.09566295 | 14 | A | 0 | 320193 | USA |
| 1690 | 1 | 20091123 | AAPL | 37833100 | APPLE INC | $ 205.88 | 2.94% | 206 | 202.95 | 203 | 3 | 1.09566295 | 14 | A | 0 | 320193 | USA |

Figure 11: CSV file snippet showing price data for AAPL.

## B Price Volatility Calculation

### B.1 News Impact Curve

The following Figure 12 shows the 'News Impact Curve' [8] used for our volatility calculation as was detailed in the methods section. The reason for including the curve for EGARCH(1,1) model was to illustrate an important **limitation** of our used model GARCH(1,1). As can be seen from the figure, the GARCH model assumes symmetric impact on volatility for both positive and negative news, and doesn't capture the asymmetric relationship, i.e. negative news having bigger impact on price volatility relative to positive - as established by numerous research literatures in the domain [French (1987), Nelson (1990)].

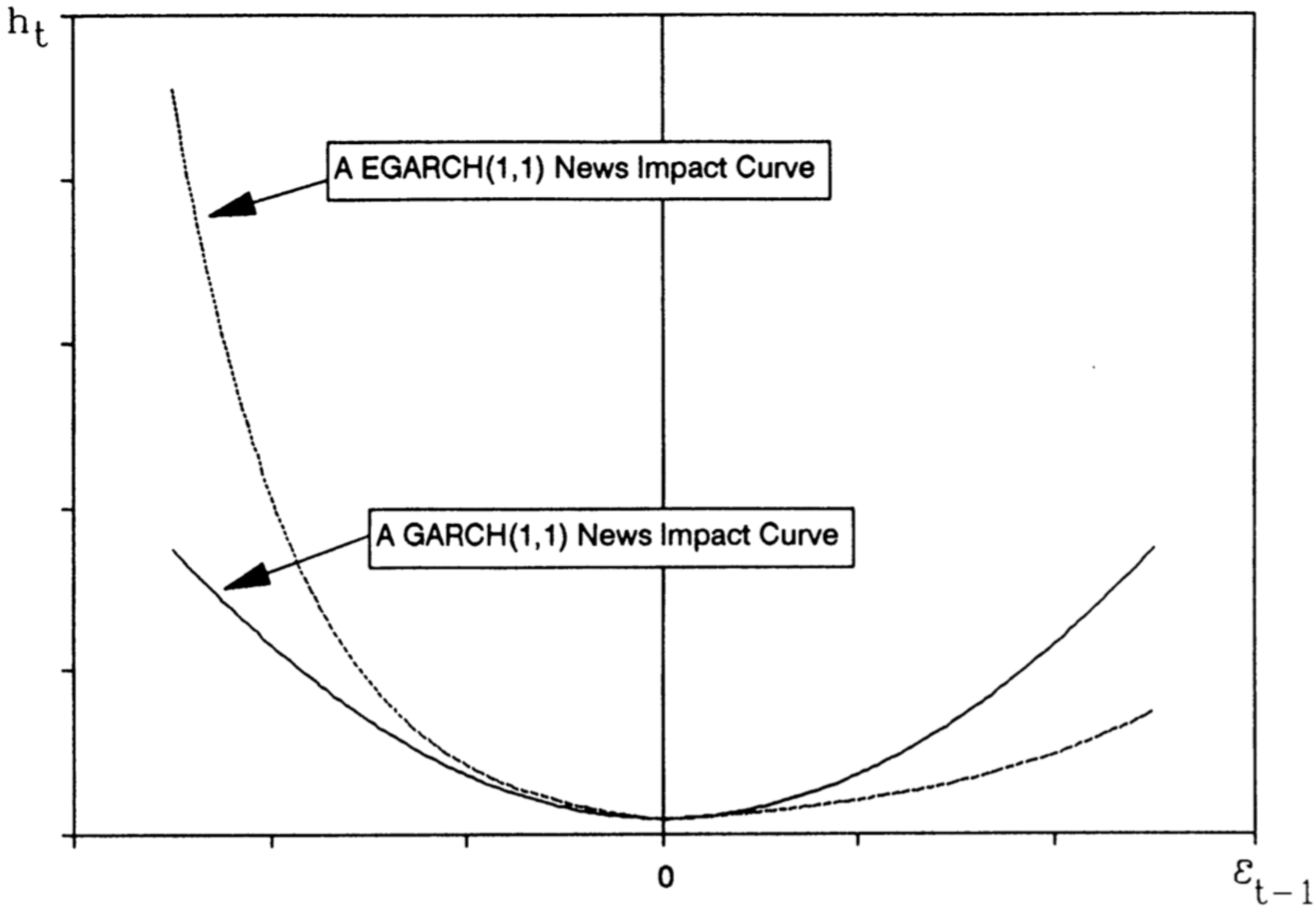

Figure 12: The news impact curves of the GARCH(1,1) model and the EGARCH(1,1) model.

### B.2 Test Accuracy for Price Volatility Calculations

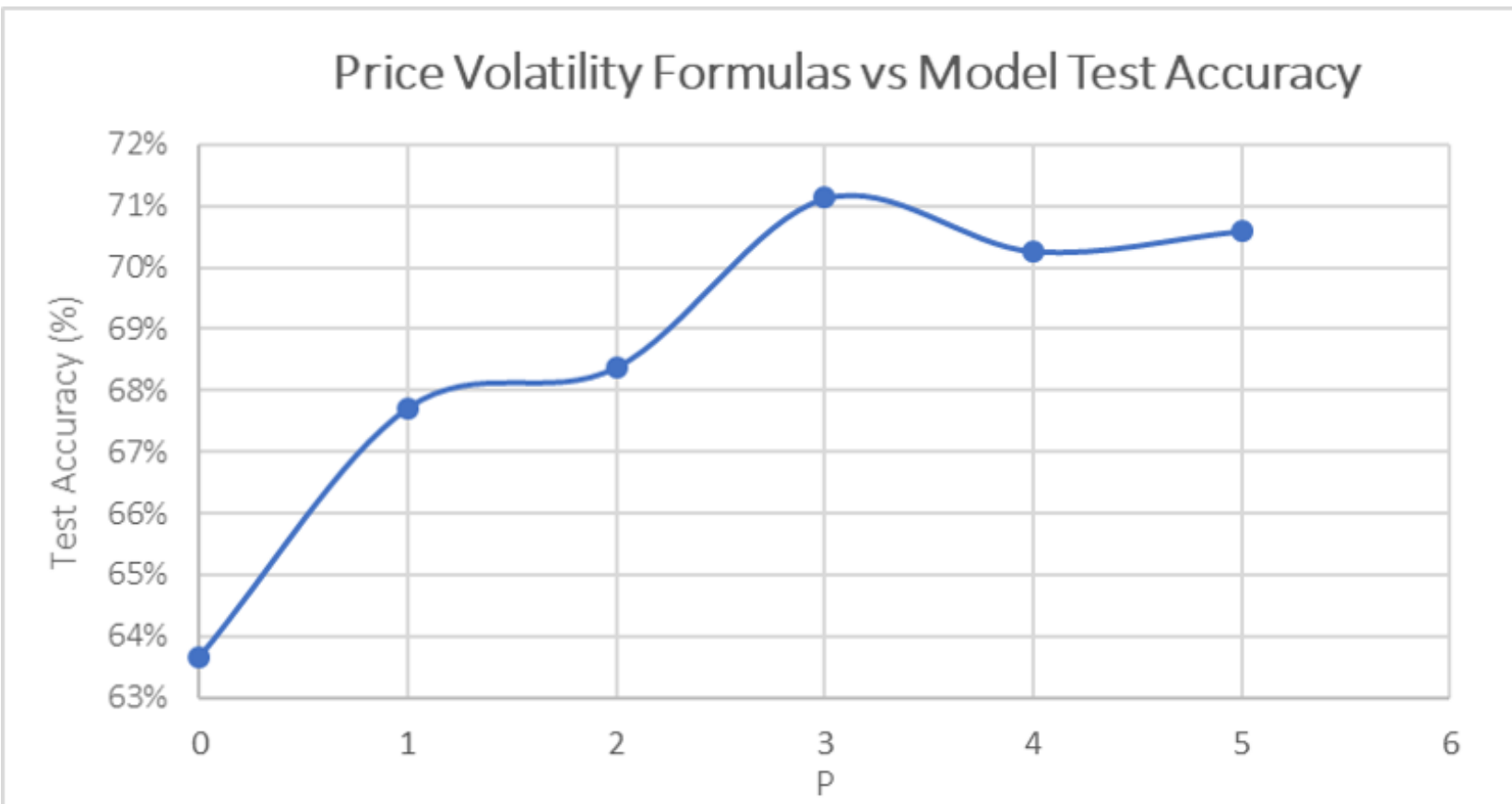

Figure 13: Test accuracy for SVM classifier with varying price volatility formula

| P | Test Error (%) |
|---|---|
| 0 | 63.67% |
| 1 | 67.72% |
| 2 | 68.38% |
| 3 | 71.13% |
| 4 | 70.27% |
| 5 | 70.60% |

Figure 14: SVM classifier test accuracy result with varying price volatility formula

## C Multi-class labeling using thresholds

In essence this is filtering based on volatility ranking, keeping the more volatile samples and discarding the lower vol samples, to allow for attribution for other exogenous market factors affecting the price movement barring news. Instead of labeling the classes binary (positive or negative) based on volatility sign, we create a neutral (safety zone) by using a 'min-threshold' and 'max-threshold' variable where the neutral class lies in the range: [min-threshold < 0 < max-threshold].

Evaluation of models using the feature selection methods mentioned in Section 4.1 were employed with machine learning techniques Bernoulli Naive Bayes and Support Vector Machines using a linear kernel. Initial evaluation was completed on 2015-2017 Reuters news data for 19 stocks from the S&P500. Price volatility was calculated using closing price from news release date and one day following. The dataset were divided by 80% and 20% train and test split.

As can be seen in Figure 15, the classifier has better precision in predicting 'positive'.

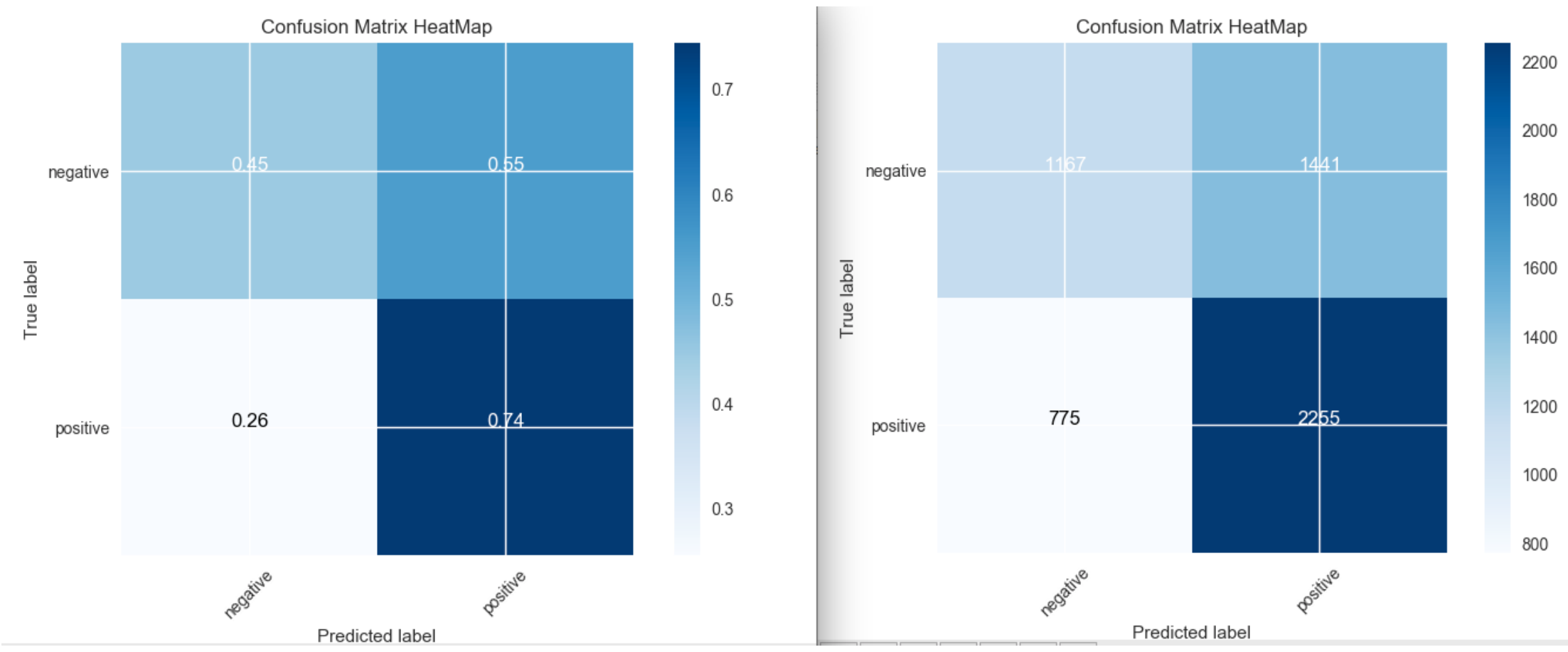

Figure 15: Confusion matrix with 2 class SVM classification with N-gram words.